\newcommand{\bee}{\begin{equation}}
\newcommand{\ene}{\end{equation}}
\newcommand{\beea}{\begin{eqnarray}}
\newcommand{\enea}{\end{eqnarray}}

\documentclass[aps,prl,showpacs]{revtex4-1}
\usepackage{graphicx}% Include figure files
\usepackage{dcolumn}% Align table columns on decimal point
\usepackage{bm}% bold math
\begin{document}
\title{Shukla-Eliasson Attractive Force: Revisited}
\author{M. Akbari-Moghanjoughi}
\affiliation{Azarbajan Shahid Madani University, Faculty of Sciences, Department of Physics, 51745-406 Tabriz, Iran}

\begin{abstract}
By investigation of the dielectric response of a Fermi-Dirac plasma in the linear limit and evaluation of the electrostatic potential around the positive stationary test charge, we find that the Shukla-Eliasson attractive force is present for the plasma density range expected in the interiors of large planets for a wide range of plasma atomic-number. This research which is based on the generalized electron Fermi-momentum further confirms the existence of the newly discovered Lennard-Jones-like attractive potential and its inevitable role in plasma crystallization in the cores of planets. Moreover, it is observed that the characteristics of the attractive potential is strongly sensitive to the variation of the plasma density and composition. Current research can also have applications in the study of strong laser-matter interactions and inertially confined plasmas.
\end{abstract}

\maketitle

\section{Introduction}

Miniaturization in technology on one hand and interest in astrophysical phenomena on the other have motivated a great deal of enthusiasm towards the study of quantum plasmas. In a series of papers Fowler, Chandrasekhar, Bohm, Pines, and Levine (\cite{fowler, chandra00, bohm, pines, levine}) have founded the concepts of relativistic-degeneracy and quantum plasma theories. Recent developments of quantum hydrodynamics (QHD) (\cite{manfredi, haas2}) and spin-1/2 quantum magnetohydrodynamics (QMHD) models (\cite{marklund0, marklund1, brodin1}) based on Wigner-Poisson and Schr\"{o}dinger-Poisson formulation (e.g. see\cite{haas0}) have opened a new area of research in the field of cold ionized plasmas. Quantum hydrodynamic theories can be used to evaluate the dynamics of spin-1/2 electron-hole systems such as semiconductors (\cite{gardner, markowich}) and nanostructured materials (\cite{shukla0}). Recent studies based on the QHD model have revealed that the particle nonlocality feature introduced into the quantum plasma model due to the Bohm force is a new feature only present in Fermi fluid models and can introduce additional critical quantum effects in the collective dynamics of Fermi-Dirac plasmas. Quantum plasmas have much reduced average interparticle distances compared to that in the classical counterparts. It has also been shown that the particle diffraction effect in the quantum plasma can lead to new multistream plasma instabilities not encountered in ordinary classical plasmas (\cite{haas3}). Extended quantum hydrodynamics formalisms have recently appeared for magnetized and unmagnetized cases taking into account the relativistic spin-orbit interaction of Fermi-Dirac particles ubiquitous in strong laser-matter interactions and astrophysical dense plasmas (\cite{mend, asenjo, haas4}).

Applications of QMHD in quantum plasmas, on the other hands, reveals that dynamics of the Fermi spin-1/2 plasma is fundamentally affected by the pressure caused due to the electron spin and orbit magnetization (\cite{marklund2, brodin2}). Furthermore, it has been shown that the magnetization in quantum plasmas can give rise to distinct degeneracy regimes (\cite{akbari2}) and remarkable behavior for spin-induced nonlinear waves (\cite{akbari3}). One of the distinguished manifestation of electron quantum diffraction effect in a degenerate plasmas is the plasma dielectric response to the external electromagnetic perturbations. The plasma dielectric susceptibility for unmagnetized and magnetized quantum plasmas, have recently been investigated by many authors. The new findings indicate the critical influence of quantum effects on the Debye shielding in Fermi plasmas \cite{saleem2}. as a fundamental problem, there has been many previous investigations on the potential around a test charge in variety of plasma environments (\cite{stan0, stan01, stan1, stan2}). More recently, by evaluating the electrodynamic potential of a moving test charge in a quantum plasma a modified Debye-H\"{u}ckel (DH) potential, i.e., the ordinary DH potential coupled with an oscillatory term have found in quantum plasmas \cite{shukla3}. Such modified DH potential, particularly caused by the presence of electron degeneracy, interaction energies and the electron nonlocality effect, has been shown to lead to a new type of Lennard-Jones-like attractive force with a pronounced potential valley near the test particle (\cite{shukla4}). The discovery of such new force in quantum plasmas has fundamental significance for the plasma coupling, crystallization and quantum transport effects in dense plasmas. The introduction of such attractive force, which brings the ions closer together, may even lead to breakthroughs in semiconductor, supercomputing and nanotechnology sciences.

The astrophysical applications of the quantum hydrostatic model became apparent, since the pioneering work of Chandrasekhar (\cite{chandra00, chandra1, chandra2}) on the relativistic degeneracy. Due to the change in the equation of state of the electron degeneracy pressure in the relativistic degeneracy limit a fundamental limit on white dwarf-mass has been discovered (\cite{chandra3}) and distinct features has been shown to exist in the QHD wave dynamics of ultrarelativistic Fermi-Dirac plasmas (\cite{akbari4}). On the other hand, the application of QMHD model suggests some large-scale effects such as modification on the well-known Jeans criteria \cite{lund} due to the presence of electron spin and orbital magnetizations. Therefore, it is appealing to examine the effect of relativistic degeneracy on quantum plasma shielding and its dielectric response. In a relativistically degenerate Fermi-Dirac plasma Salpeter (\cite{salpeter}) has evaluated the effects of variety of interaction potentials such as Coulomb lattice-energy, Thomas-Fermi distribution, electron-exchange, and correlation on stellar hydrostatic stability for a wide range of relativity parameter and has found more significant contribution due to Coulomb lattice energy compared to others interaction effects. Such a plasma interaction effect although being a few percent of the dominant electron degeneracy effect, can give rise to crystallization of the Fermi-Dirac plasma in superdense stellar cores. In this paper we aim at examining the linear plasma response and the potential of a test charge due to the presence of Coulomb interactions and the Bohm force in relativistically degenerate Fermi-Dirac plasmas for a wide range of relativity parameter (electron number-density). We present the quantum hydrodynamics (QHD) model in Sec. \ref{Model} and examine the plasma dielectric constant in the linear limit in Sec. \ref{Response}. Then, we calculate the modified Debye-Huckel potential around a stationary positive test charge in Sec. \ref{Crystal} giving the parametric description of the associated potential in Sec. \ref{Numerical}. Finally, we make numerical evaluation of the calculated potential based on independent plasma parameters in Sec. \ref{Discussion} and draw our conclusions in Sec. \ref{conclusion}.

\section{Quantum Hydrodynamics Model}\label{Model}

In this model we consider a plasma of nondegenerate stationary ions with charge $Z$ quasineutrally distributed into a completely degenerate Fermi-Dirac electron fluid. The ion thermal energy is neglected compared to the electron Fermi-energy ($E_{Fe}$), hence, they are assumed to be static and unperturbed to external simulations compared to light electrons ($m_e/m_i\ll 1$). Furthermore, we use the nearly-free electron model neglecting the electron fluid viscosity. The electron viscosity has been shown to becomes dominant in much larger densities such as in neutron-star crusts \cite{itoh}. On the other hand, it is also that in quantum plasmas the electron-electron collisions are limited due to the Pauli blocking-mechanism (\cite{shukla0}). In the electron number-density regime, considered in this work ($10^{22}-10^{26}/cm^3$), one may also assume infinite plasma conductivity ($\sigma\simeq 10^{22}$) thereby neglecting the ion-electron collision frequency (\cite{radha}) compared with the characteristic plasma frequencies. Therefore, the hydrodynamic evolution of Fermi-Dirac electron gas is governed by the following set of equations (\cite{crus}), as follows
\begin{equation}\label{dimensional}
\begin{array}{l}
\frac{{\partial n}}{{\partial t}} + \nabla \cdot(n{\bf{u}}) = 0, \\
\frac{{\partial {\bf{u}}}}{{\partial t}} + ({\bf{u}}\cdot\nabla ){\bf{u}} = e\nabla \phi  - \frac{1}{{{m_e}n}}\nabla {P_G} + \frac{{{\hbar ^2}}}{{2m_e^2}}\nabla \left[ {\Delta \sqrt n /\sqrt n } \right], \\
\Delta \phi ({\bf{r}}) = 4\pi e(n - {n_0}), \\
\end{array}
\end{equation}
where, $\hbar$ and $c$ are scaled Plank-constant and speed of light in vacuum, respectively. Also, $P_G$ is the generalized pressure of plasma composed of electron degeneracy, exchange and Coulomb interaction pressures, i.e., $P_G=P_d+P_{xc}+P_C$ and $n_0$ is the unperturbed electron density. Ignoring the negligible ion pressure compared to that of the quantum pressure of electrons, a general form of electron degeneracy pressure for a wide range of plasma density is given as (\cite{chandra1})
\begin{equation}\label{pd}
{P_e} = \frac{{\pi m_e^4{c^5}}}{{3{h^3}}}\left\{ {R\left( {2{R^2} - 3} \right)\sqrt {1 + {R^2}}  + 3\ln \left[ {R + \sqrt {1 + {R^2}} } \right]} \right\},
\end{equation}
where, the Chandrasekhar relativity parameter, $R$, is related to the Fermi relativistic-momentum via the relation; $R=p_{Fe}/m_e c=(n/n_c)^{1/3}$ with $n_c\simeq 5.9\times 10^{29} cm^{-3}$ being the normalizing value. Furthermore, a relativistic degeneracy parameter, $R_0$, may be introduced as $R=R_0(n/n_0)^{1/3}$ where $R_0=(n_0/n_c)^{1/3}$ with $n_0$ being the unperturbed equilibrium electron number-density. It can be easily shown that, for the degenerate electron pressure represented as a polytropic form of $P=P_0(n/n_0)^\gamma$ the values of $\gamma=\{5/3,4/3\}$ correspond to the degeneracy limits $R=\{0,\infty\}$ of Eq. (\ref{pd}) as nonrelativistic and ultrarelativistic degeneracy limits, respectively. The change in equation of state due to the relativistic degeneracy of electrons has been shown to lead to the well-known white-dwarf mass-limit (\cite{chandra2}) and distinct features in nonlinear wave dynamics in Fermi-Dirac plasmas (\cite{akbari2}). It can be observed that the polytropic form of the Chanrasekhar pressure approaches nicely to the standard Fermi degeneracy pressure for the extremely nonrelativistic number-density range \cite{akbari6}. It is evident that one needs to use the full form of pressure, valid for a wide plasma density-range, in order to evaluate the Shukla-Eliasson force and phase separation in dense quantum plasma to be shown present also in the interiors of large planets.

As the plasma becomes denser, the corrections to the equation of state of the plasma due to the particle interactions become important. Such interaction terms in the strong coupling plasma limit are due to the electrostatic plasma interactions (Coulomb interactions), electron exchange, electron nonuniform distribution (Thomas-Fermi correction), ion correlation and some other minor effects (\cite{salpeter}). It is known that the plasma coupling parameter, $\Gamma$, is inversely proportional to the interparticle distances, $r_{ij}$, and the temperature $T$ of the plasma species. Hence, with the interparticle distances being of the same order for ions and electrons in the quasineutrality limit, the coupling parameter for ions is observed to be much larger than that for degenerate electrons, since, the interacting electron temperature is of the order of the Fermi-temperature being much higher than the ion thermal temperature ($\Gamma_i\gg\Gamma_e$). The Coulomb attractive potential, being the dominant interaction potential, is the main cause of the crystallization of Fermi-Dirac plasmas, under strong coupling conditions. Its dominant effects is well-understood in ordinary solids-state matter and is believed to play a fundamental role in Fermi-Dirac plasmas lattice formation. Calculation of the collective electrostatic energy for crystals with different periodic configurations such as the \emph{fcc} and the \emph{hcp} lattice structures confirm that the Coulomb energy of a lattice is minimally related to the lattice structure (\cite{kittel}). Salpeter (\cite{salpeter}), using a spherical noninteracting Wigner-Seitz cell of radius, $r_0$ ($1/n_0=4\pi r_0^3/3$), has given an expression for the Coulomb energy of a relativistically degenerate plasma with ions forming a quasi-lattice at very high electron-density plasma regime such as in a white dwarf core or neutron star crust. It has been shown that such attractive force persists and even becomes stronger as the value of the relativity parameter or the electron number-density increases. On the other hand, the plasma coupling parameter which is given by $\Gamma=(Ze)^2/(r_0 k_B T)$ may also be approximated in terms of the relativity parameter as; $\Gamma  \simeq 0.23{Z^{5/3}}R/T_8$, (where $T_8$ is the plasma temperature in units of $10^8$ degrees Kelvin) (\cite{itoh}). It is known that, the plasma is in liquid form for the coupling parameter range $\Gamma<178$ and in solid phase beyond this value (\cite{itoh}). Hence, it is concluded that for a typical planet core temperature, say $24000K$ for Jupiter core, and a typical relativity parameter of $R\simeq0.1$, the coupling parameter value exceeds the liquid critical limiting value. Therefore, it is observed that, the planet cores for a wide range of plasma atomic-number value would be in a quasi-lattice phase. The effects of interaction potential on ion wave dynamics in Fermi-Dirac plasmas has been investigated recently \cite{akbari7}.
\begin{equation}\label{ctf}
{P_C} =  - \frac{{8{\pi ^3}m_e^4{c^5}}}{{{h^3}}}\left[ {\frac{{a_0 {Z^{2/3}}}}{{10{\pi ^2}}}{{\left( {\frac{4}{{9\pi }}} \right)}^{1/3}}} \right]{R^4}
\end{equation}
where, $Z$ is the atomic number of ions and $h$ is the Planck constant. It should be noted that, the Coulomb pressure presented here only gives a qualitatively right order for $Z^{2/3}(r_0 /r_B)\simeq 1$ in the nonrelativistic plasma density limit ($10^{25}<n_0/cm^{3}<10^{27}$), where, $r_0$ and $r_B$ are the Wigner-Seitz and Bohr radiuses, respectively. Therefore, in the proceeding we will confine our analysis to this validity limit which is relevant for the planet core density range. On the other hand, the contribution from the electron exchange may be written as,
\begin{equation}\label{ex}
\begin{array}{l}
{P_{xc}} =  - \frac{{2a_0 m_e^4{c^5}}}{{{h^3}}}\left\{ {\frac{1}{{32}}\left( {{\beta ^4} + {\beta ^{ - 4}}} \right) + \frac{1}{4}\left( {{\beta ^2} + {\beta ^{ - 2}}} \right) - \frac{3}{4}\left( {{\beta ^2} - {\beta ^{ - 2}}} \right)\ln \beta } \right. + \frac{3}{2}{\left( {\ln \beta } \right)^2} \\ - \frac{9}{{16}}
\left. { - \frac{R}{3}\left( {1 + \frac{R}{{\sqrt {1 + {R^2}} }}} \right)\left[ {\frac{1}{8}\left( {{\beta ^3} - {\beta ^{ - 5}}} \right) - \frac{1}{4}\left( {\beta  - {\beta ^{ - 3}}} \right) - \frac{3}{2}\left( {\beta  + {\beta ^{ - 3}}} \right)\ln \beta  + \frac{{3\ln \beta }}{\beta }} \right]} \right\}. \\
\end{array}
\end{equation}
where, $\beta = R + \sqrt{1 + R^2}$. The Coulomb interaction pressure is always negative, while, the pressure due to electron exchange changes the sign at the relativity parameter value around $R\simeq 2$ \cite{akbari7}. It should be noted that in the forthcoming calculations we have included the generalized form of the exchange energy, given by Salpeter, instead of that in (\cite{shukla4}), since, the generalized exchange energy is valid for a wide electron number-density range. We further have neglected the effect of electron-correlation effect due to the vanishingly small contribution in the considered density regime (\cite{salpeter}).

\section{Linear Dielectric Response}\label{Response}

In order to obtain the dielectric constant for the present plasma model we use a standard procedure employed by Shukla \& Eliasson (\cite{shukla4}). The QHD equations given in Eqs. (\ref{dimensional}) now read as
\begin{equation}\label{norm}
\begin{array}{l}
\frac{{\partial n}}{{\partial t}} + \nabla \cdot n{\bf{u}} = 0, \\
\frac{{\partial {\bf{u}}}}{{\partial t}} + ({\bf{u}}\cdot\nabla ){\bf{u}} = \frac{e}{{{m_e}}}\nabla \phi  - {c^2}\nabla \left[ {{\Psi _d} + {\Psi _{xc}} + {\Psi _C}} \right] + \frac{{{\hbar ^2}}}{{2m_e^2}}\nabla \frac{{\Delta \sqrt n }}{{\sqrt n }}, \\
\Delta \phi ({\bf{r}}) = 4\pi e(n - {n_0}), \\
\end{array}
\end{equation}
where, $\Psi$'s are the effective plasma potentials defined via the relation
\begin{equation}\label{dimensional}
\Psi  = \frac{1}{m_e}\int {\frac{{{d_R}P(R)}}{R^3 }} dR,\hspace{3mm}\nabla {\Psi} = \frac{1}{m_e}\frac{1}{R^3 }\frac{{d{P}(R)}}{{dR}}\nabla R.
\end{equation}
The plasma effective potentials are, then, given as
\begin{equation}\label{eff}
{\Psi _d} = \sqrt {1 + R_0^2{n^{2/3}}},\hspace{3mm}{\Psi _{xc}} = \frac{{{a_0}}}{{2\pi }}\left[ {{R_0}{n^{1/3}} - \frac{{3{{\sinh }^{ - 1}}({R_0}{n^{1/3}})}}{{\sqrt {1 + R_0^2{n^{2/3}}} }}} \right],\hspace{3mm}{\Psi _C} =  - \beta {R_0}{n^{1/3}}, \\
\end{equation}
with
\begin{equation}\label{beta}
\beta  = \frac{{2{a_0}}}{5}{\left( {\frac{{2Z}}{{3\sqrt \pi  }}} \right)^{2/3}},\hspace{3mm}a_0\simeq 1/137.
\end{equation}
To evaluate the linear response of the plasma to harmonic excitations of form $\propto \exp [i({\bf{k}} \cdot {\bf{r}} - \omega {\rm{t}})]$, we use the well-known harmonic operators $\nabla\equiv i\bf{k}$ and $\partial_{t}\equiv -i\omega$ to linearize the equation set Eqs. (\ref{norm}). Assuming a small density perturbation around the equilibrium state (i.e. $\bf{u}=\bf{u_1}$, $\phi=\phi_1$, and $n=n_0+n_1$ with $n_1\ll n_0$), we obtain the following equations
\begin{equation}\label{eff}
\begin{array}{l}
- \omega {n_1} + {\bf{k}}\cdot{{\bf{u}}_{\bf{1}}} = {\rm{0}}, \\
- \omega {{\bf{u}}_{\bf{1}}} =  - \frac{{{\hbar ^2}}}{{4m_e^2}}{k^2}{\bf{k}}{n_1} - {c^2}T{\bf{k}}{n_1},\\
T = \frac{{{R_0^2}}}{{3\sqrt {1 + {R_0^2}} }} - \frac{\beta}{3} R_0 + \frac{a_0 }{{2\pi }}\left[ {\frac{R_0}{3} - \frac{R_0}{{1 + {R_0^2}}} + \frac{{{R_0^2}{{\sinh }^{ - 1}}R_0}}{{{{(1 + {R_0^2})}^{3/2}}}}} \right],
\end{array}
\end{equation}
the reduction of which yields the first order potential perturbation and the electron dielectric susceptibility, as
\begin{equation}\label{disp}
\chi _e^{ - 1} =  - \frac{{{k^2}{\phi _1}}}{{{n_1}}} = \frac{{{\hbar ^2}}}{{4m_e^2}}{k^4} + {c^2}T - {\omega ^2},
\end{equation}
where, $\chi_e$ is the dielectric susceptibility related to the dielectric constant as $\epsilon(\omega,\bf{k})=\rm{1}+\chi_e$. In the quasistationary limit, we obtain the normalized dielectric constant, as  (\cite{shukla4})
\begin{equation}\label{di}
\epsilon(0,{\bf{k}}) = 1 + \omega _{pe}^2{\left[ {\frac{{{\hbar ^2}}}{{4m_e^2}}{k^4} + {c^2}T} \right]^{ - 1}},
\end{equation}
where, $\omega_{pe}=\sqrt{4\pi n_0 e^2/m_e}$ is the electron plasmon frequency. Note that, in many recent quantum plasma literature the use of the quantum diffraction parameter, $H=\hbar^2 \omega_{pi}^2/(2 m_e c^2)$ called the scaled ion-plasmon frequency, has become widespread. The value of this parameter can be approximated as $H\simeq0.0044ZR_0^{3/2}$ which for metallic iron density of $\simeq 10^{23}/cm^3$ and higher density of our concern ($10^{26}/cm^3$) with same composition take the values of $H\simeq 0.00005$ and $H\simeq 0.002$, respectively. However, the corresponding values are much smaller than that suggested previously (\cite{haas2}) for metallic solids which is due to the normalization factor used in this work.

\section{Phase Separation in Fermi-Dirac Plasmas}\label{Crystal}

Let us consider a positive stationary test charge in the plasma with Fermi-Dirac degenerate electrons. The shielding of such a test charge is a fundamental property of the plasma and can give useful information about the bound states between ions. The Debye-shielding or the near-field electrostatic potential usually appears as $\phi(r)\propto(1/r)exp(-r/r_d)$ where $r_d$ is the Debye-length and $r$ is the distance from the test charge (\cite{krall}). In the extreme nonrelativistic degeneracy limit the plasma shielding potential can attribute to also a far-field inverse cubic dependence ($r^{-3}$) from an outside inertial observer point of view (\cite{shukla3}). Let us for simplicity assume a stationary test charge $Q_T=Q$ located at the origin. It is obvious that analysis can be extended to a moving test charge with velocity $\bf{v}_{\rm{T}}$ relative to the laboratory frame (\cite{shukla3}). The Poisson equation in the presence of point-like stationary positive test charge can be written as,
\begin{equation}\label{poi}
\Delta \phi ({\bf{r}}) = 4\pi e(n - {n_0}) - 4\pi Q\delta ({\bf{r}}),
\end{equation}
where, $\delta ({\bf{r}})$ is the Dirac delta-function ensuring the localization of the test charge at the origin, i.e., $\delta ({\bf{0}})=1$ and $\delta ({\bf{r\ne 0}})=0$. Using the Fourier analysis in space, the electrostatic potential around the test charge can be rewritten in the form
\begin{equation}\label{phi}
\phi ({\bf{r}}) = {\frac{Q}{{2\pi^2 }}}\int {\frac{{\exp (i{\bf{k}} \cdot {\bf{r}})}}{{{k^2}\epsilon({\bf{k}})}}{d^3}{{k}}},
\end{equation}
where, $\epsilon(\bf{k})$ is the plasma dielectric constant, given by Eq. (\ref{disp}). We then appropriately decompose the integral following the procedure given by \cite{shukla4}. Therefore, the integral in Eq. (\ref{phi}) can be decomposed into different parts, as follows
\begin{equation}\label{dec}
\phi ({\bf{r}}) = {\frac{Q}{{4\pi^2 }}}\int {\left[ {\frac{{1 + L }}{{{k^2} + k_1^2}} + \frac{{1 - L }}{{{k^2} + k_2^2}}} \right]\exp (i{\bf{k}} \cdot {\bf{r}}){d^3}k},
\end{equation}
where, the parameters $k_{1,2}$ and $L$ are given as
\begin{equation}\label{ks}
k_{1,2}^2 = k_0^2\frac{{1 \mp \sqrt {1 - 4\alpha } }}{{2\alpha }},\hspace{3mm} L = \sqrt {1 - 4\alpha } . \\
\end{equation}
with the parameters $k_0$ and $\alpha$ given as
\begin{equation}\label{pars}
\alpha  = \frac{{{\hbar ^2}\omega _p^2}}{{4{m_e^2}c^4{T^2}}},\hspace{3mm}{k_0} = \frac{{{\omega _p}}}{{c\sqrt T }}
\end{equation}
Now, using the condition, $\phi(r=\infty)=0$, and the standard definition
\begin{equation}\label{id}
\int {\frac{{\exp (i{\bf{k}}\cdot{\bf{r}})}}{{{k^2} + k_{1,2}^2}}{d^3}k} = \frac{{2{\pi ^2}}}{r}{e^{ - {k_{1,2}}r}},
\end{equation}
we may write the general solution for the first-order potential perturbation around the test charge as,
\begin{equation}\label{sol}
\phi (r) = \frac{Q}{{2 r}}\left[ {\left( {1 + L } \right){e^{ - {k_1}r}} + \left( {1 - L } \right){e^{ - {k_2}r}}} \right].
\end{equation}
We, then, are in a position to evaluate the potential around the test charge in the Fermi-Dirac degenerate plasma with respect to two independent plasma parameters, namely, the relativistic degeneracy parameter, $R_0$, representing the normalized number-density of electrons and the atomic-number, $Z$, of the plasma ions. As it is evident, this can be only done in a numerical scheme presented below. We will present the phase separation in Fermi-Dirac plasma for different values of the plasma atomic-number using the scaled Bohr-radius, $r_s=r_B/r_0$, where, $r_B=\hbar^2/m_e e^2$ is the Bohr-radius. This implies the relation; ${R_0} \approx 0.014{r_s}$ between the degeneracy parameter, $R_0$, and the scaled Bohr-radius, $r_s$.

\section{Shukla-Eliasson Attractive Potential in Fermi-Dirac Plasmas}\label{Numerical}

It is evident that for a physical solution to exist, in Eq. (\ref{sol}), the value of the potential must be real. It can be easily confirmed that for all values of $\alpha>1/4$ the parameter $L$ is imaginary. Therefore, for $\alpha> 1/4$ (it can be confirmed that the parameter $\alpha$ is always positive), we have $\Re{(k_1)}=\Re{(k_2)}$ and $\Im{(k_1)}=-\Im{(k_2)}$, hence, $k_{1,2}=\Re{(k_{1})}\pm \Im{(k_{1})}$ (\cite{shukla4}) and the electrostatic potential can be written in the form $\phi (r)=(Q/r) \exp[ - \Re ({k_2})r]\{ \cos [\Re({k_2})r] + \Re{(L)}\sin [\Im ({k_2})r]\}$, where $\Re$ and $\Im$ refer to the real and imaginary parts of the parameter $k_2$. This attractive Lennard-Jones-like potential relates to the ion bound-states of the Fermi-Dirac plasma. However, for $\alpha<1/4$, where the parameter $L$ becomes real-valued, the potential of a test charge reduces to the repulsive Debye-Huckel or Yukawa-type potential. It is important to note that the Shukla-Eliasson force is an elegant manifestation of pure quantum mechanical effect and a consequence of the electron collective quantum behavior, since, it disappears in the $\hbar\rightarrow 0$ limit. It is also observed that, the characteristics of such attractive force depends strongly on the atomic-number of the plasma ions. As it will be revealed in Sec. \ref{Discussion}, our research highlights the importance of quantum tunneling effect on electrons through the high plasma number-density and the Coulomb negative pressure. It is also observed from Eq. (\ref{pars}) that below a critical electron number-density, $n_{0cr}$, for a given atomic-number, $Z$, the parameter $T$ can becomes negative leading to an imaginary value for $k_0$ and consequently for the screening potential. Therefore, there is a lower plasma number-density bound, where the attractive force can exist. This is completely natural, since, as the plasma density is lowered the degeneracy limit of Fermi-Dirac plasma is reached beyond which the attractive force should disappear.

It is most surprising, however, to see some recent literature (\cite{vranjes}) to question such rigorously furnished theory as quantum plasma theory and its fundamental applications, based on fallacious arguments. Also, based on the density functional theory computer simulations some authors (\cite{bonitz}) have argued against the applicability of linearized quantum hydrodynamic theory and the existence of Shukla-Eliasson attractive potential. One must note that, the development of improved functional theory is currently a very active area of research and it is far from clear whether the research will be successful in providing the substantial increase in accuracy desired (\cite{sch}). In the following numerical analysis it is confirmed that such attractive potential not only persists in high density matter it also may be one of the essential ingredients causing the strong ion-coupling in the cores of planets.

\section{Numerical Analysis of Phase-Separation}\label{Discussion}

\begin{figure}[ptb]\label{Figure1}
\includegraphics[scale=.5]{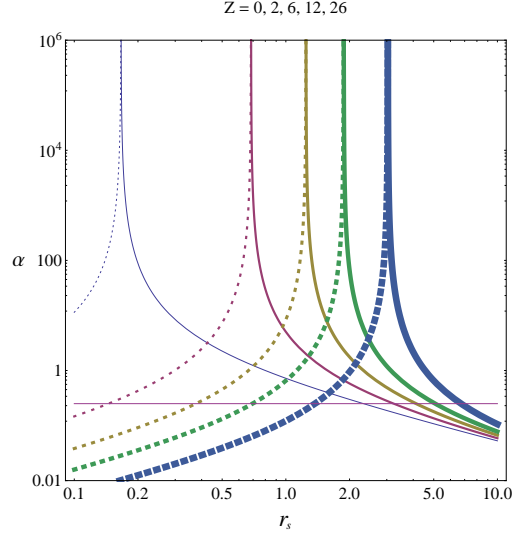}\caption{Fig. 1 shows the phase-separation diagram of the Fermi-Dirac plasma without ($Z=0$) and with Coulomb correction ($Z\neq0$) in Log-Log scale. The horizontal line in figure indicates the limit above(below) which the potential is attractive(repulsive). The solid/dashed branches show the imaginary/real parts of the phase diagram. The thickness of curves was used as a scale to show the variation in the plasma atomic-number.}
\end{figure}
Figure 1 shows the phase separation parameter, $\alpha$, for the extended quantum plasma model in terms of the scaled fermion spacing parameter, $r_s$ for different values of the plasma atomic-number. The figure consists of phase-separation curves for different plasma composition and each curve has two branches. It is observed that, for the dashed branch the parameter, $T$ becomes negative and the potential imaginary. The solid branch, on the other hand, gives rise to the real electrostatic potential. The part of the phase-separation curve above the horizontal line indicates the presence of the Shukla-Eliasson attractive force and part below the horizontal line corresponds to the ordinary Debye screening-potential. It revealed that as the plasma density increases ($r_s$ increases) at some point, depending on the plasma atomic-number, the attractive potential appears and by further increase in the plasma density (increase in $r_s$) at some point, depending on the plasma atomic-number, the attractive potential changes to the repulsive one. Note that the at the critical point, $\alpha=\infty$, where attractive force is about to appear, the force between ions is the ordinary Coulomb-force as it is expected. Also at the point $\alpha=1/4$, where the attractive force is about to disappear, the potential is finite and given in Ref. (\cite{shukla4}). It is remarked from Fig. 1 that, the starting and ending points for the attractive potential is strongly dependent to the plasma composition and moves to larger densities for larger atomic-number. This means that, the attractive force in quantum plasmas starts and ends at higher plasma densities for higher atomic-number plasmas. It is also revealed that, density range for which the attractive Shukla-Eliasson force exists narrows as the plasma atomic-number increases. It is critical to note that the Shukla-Eliasson force for iron composition starts at approximate density of $\rho\simeq 75.6675 gr/cm^3$ which is much higher than the zero-pressure solid iron with density around $8 gr/cm^3$. This is an indication of the fact that the binding mechanism of Shukla-Eliasson force which is due to the electron tunneling and exchange interaction effects is quite different from that of the ordinary solids.
\begin{figure}[ptb]\label{Figure2}
\includegraphics[scale=.46]{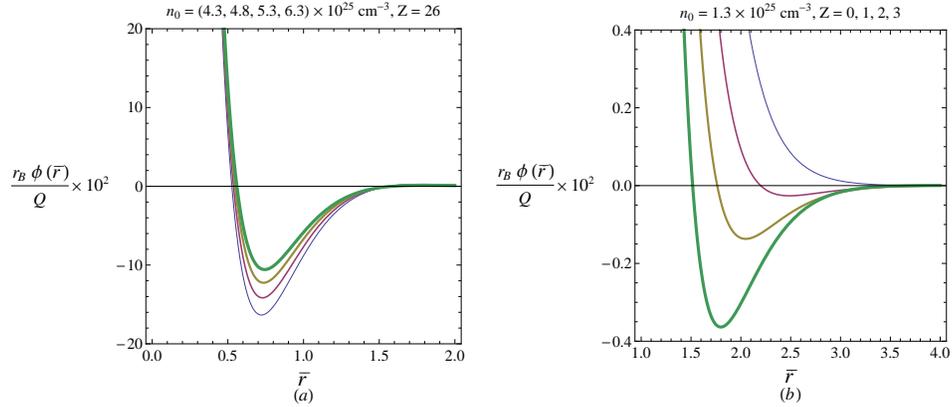}\caption{Fig. 2 shows the variation of Shukla-Eliasson attractive potential depth and scaled position with respect to change in the plasma number density (plot 2(a)) and atomic number (plot 2(b)) in the low density plasma limit. The thin curves in each plot corresponds to the $Z=0$ case and the thickness of curves was used as a scale to show the variation in the plasma atomic-number and the plasma number-density.}
\end{figure}
Figure 2 shows the characteristics of the attractive potential for various plasma parameters. It is remarked from Fig. 2(a) that the increase in the plasma number-density leads to the weakening of the attractive force for a given plasma number-density ($Z=26$ for iron composition). As it was noted in above paragraph, the attractive force for iron-composed plasma happens at much larger densities than that of the ordinary solid iron (e. g. see Fig. 1(a)). It is also remarked from Fig. 1(a) that the ion separation in Shukla-Eliasson quasi-lattice is almost independent of the plasma number-density variation. Figure 1(b) shows the variation in potential profile parameters (the binding energy and the quasi-lattice parameter) due to change in the plasma composition parameter, $Z$. It is clearly observed that both the binding energy and the lattice parameters are altered significantly due to the change in the plasma atomic-number.
\begin{figure}[ptb]\label{Figure3}
\includegraphics[scale=.47]{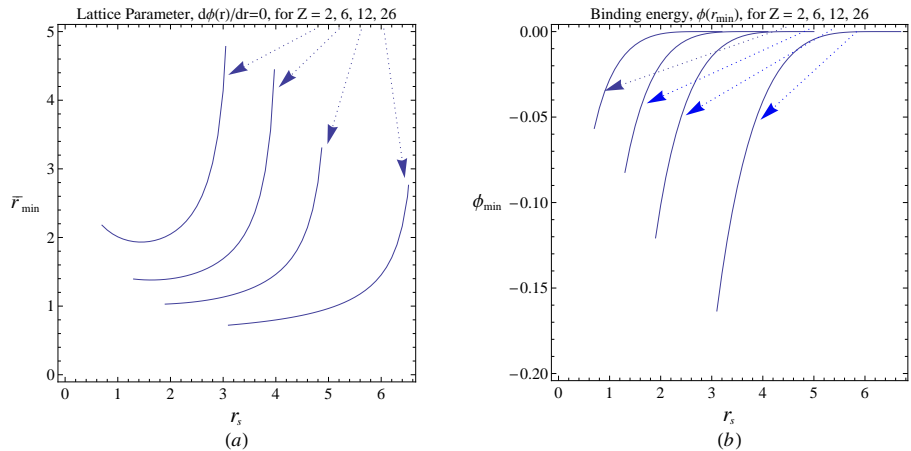}\caption{Fig. 3 shows the regions and variation of the attractive potential parameters, i.e., the position of the potential minimum, $\bar r_{min}$ (Fig. 3(a)), and the binding energy, $\phi(\bar r_{min})$ Fig. 3(b), for different values of the atomic-number, $Z$, in terms of the reduced Wigner-Seitz redius, $r_s$.}
\end{figure}
From Fig. 3(a) it is observed that the equilibrium distance, $\bar r_{min}$, and the plasma binding energy (the depth of the potential minimum, $\phi({\bar r_{min}})$) increases strongly as the plasma-density (or the reduced Wigner-Seitz radius) increases for all values of the atomic-number, $Z$. Figure 3(b) further reveals that the plasma binding-energy (the depth of the potential minimum, $\phi({\bar r_{min}})$) decreases strongly as the plasma-density (or the reduced Wigner-Seitz radius) increases for all values of the atomic-number, $Z$. However, it is remarked that the attractive potential occurs at higher densities for higher atomic-number values. This can be an indication of the likelihood of presence of heavier compounds at the cores of large planets like Jupiter in our solar system. Figure 3 also confirms that there are two (higher and lower) cutoff plasma density values for every given atomic-number value which limit the occurrence of the Shukla-Eliasson attractive force. It also observed that the attractive force starts closer to the test charge and with higher binding-energy values for higher atomic-number of plasmas. The increase of the plasma number-density is also observed to decrease the ion separation leading to formation of more compact with higher melting point quasi-lattices. Current research highlights features in super-dense degenerate plasma which can be of vital importance in the study of planetary sciences.

\section{Concluding Remarks}\label{conclusion}

We used the linear dielectric response of a dense degenerate plasma of electron number-density in the range of that in the cores of big planets to show that the Shukla-Eliasson attractive force exists in such density regimes. It was found that the nature of newly discovered attractive force in degenerate plasmas is quite different from those in ordinary solids and the binding length and strength in this density regime significantly depends on the plasma density and composition. It was observed that heavier elements in the plasma give rise to stronger ion-binding. It was also remarked that the heavier the plasma ions the denser should the plasma be in order for the plasma binding to occur. This research highlights features which can be of fundamental importance in the study of the planetary core structures as well as inertial confined laboratory plasmas.

\acknowledgments
The author expresses his deepest gratitude to Prof. Padma Kant Shukla for his generosity in exchanging his precious ideas. Insightful comments of anonymous referees which have led the paper to its current state are also truly appreciated.

\end{document}